# Polaron and Strain Effects on Ion Migration in WO$_3$

**Matthäus Siebenhofer[1,#], Pjotrs Žguns[2] and Bilge Yildiz[1,2*]**

[1] Department of Nuclear Engineering, Massachusetts Institute of Technology, Cambridge, USA
[2] Department of Materials Science and Engineering, Massachusetts Institute of Technology, Cambridge, USA

* byildiz@mit.edu
# Current Address: Institute of Chemical Technologies and Analytics, TU Wien, Vienna, Austria

## Abstract

Ion migration in WO$_3$ is a critical process for various technological applications, such as in batteries, electrochromic devices and energy-efficient brain-inspired computing devices. In this study, we investigate the migration mechanisms of H$^+$, Li$^+$, and Mg$^{2+}$ ions in monoclinic WO$_3$, and how energy barriers are affected by the presence of electron polarons and by lattice strain. Our approach in calculating the migration paths and barriers is based on density functional theory methods. The results show that the presence of polarons leads to association effects and lattice deformations that increase ion migration barriers. Therefore, the consideration of polarons is critical to accurately predict activation energies of ion migration. We further show that lattice strain modulates ion migration barriers, however, the impact of strain depends on the migrating ion. For protons that are embedded in the oxygen ion electronic shells and hop from donor to acceptor oxygens, compressive lattice strain accelerates migration by reducing the donor-acceptor distance. In contrast, the migration barriers of larger ions decrease with tensile lattice strain that increases the free space for the ion in the transition state. These insights into the effects of polarons and lattice strain are important for understanding and tuning properties of WO$_3$ when aiming for optimized device characteristics.

## 1. Introduction

Tungsten trioxide (WO$_3$) is a semiconducting transition metal oxide that is used in a range of technological applications. With electrically tunable optical properties, it has been used in electrochromic devices for decades[1–3]. More recently, WO$_3$ was identified as a promising channel material in electrochemical random-access memory (ECRAM) devices[4–6] for low energy brain-inspired computing. Like electrochromic devices, ECRAM devices rely on the intercalation of ions, such as H$^+$, Li$^+$, or Mg$^{2+}$ into the tungsten oxide lattice [7]. However, while electrochromic devices exploit the changes of optical constants upon intercalation, the ECRAM functionality is facilitated by modulation of its electronic conductivity as a function of ion concentration[6]. For both, fast ion migration in WO$_3$ is important for reaching the desired device operation speed[4], and strategies to accelerate ion migration are of high technological relevance.

While ion migration in WO$_3$ has been studied both experimentally[8,9] and computationally[10–12], calculated migration barriers and experimental activation energies of diffusion scatter significantly and often do not align. In the case of H$^+$ diffusion, experimentally, activation energies in polycrystalline samples were found to be close to 0.5 eV[9,13]. Lin et al. computationally determined a proton transfer barrier of 0.35 eV[12]. Xi et al. found a rather low value of 0.23 eV for the highest migration barrier on a specific migration path[10]. Cui et al. found barriers between 0.13 and 1.62 eV for different migration directions[11]. For Li$^+$ migration, experiments for amorphous WO$_3$ in the relevant lithiation regime suggest a barrier of around 0.5 eV[14,15]. Previously calculated Li$^+$ migration barriers range between 0.15 eV[11,16] and 0.57 eV[15]. For Mg$^{2+}$, ion migration has not been studied experimentally or computationally yet. We suggest two potential reasons for these discrepancies, next to differences in experimental and computational methodology. One is the presence of polaronic charge carriers at low ion concentrations[17,18], that may alter ion migration dynamics. While experiments consider the real material, where a variety of electronic and ionic defects, as well as microstructure defects, affect charge transport, computational studies typically study one ion or one type of electronic defect at a time, neglecting potentially critical interactions among different charge carriers. In addition, lattice disorder and strain can be introduced in experiment, especially during thin-film deposition. Elastic strain is for example known to tune ion migration dynamics in oxide ion conductors [19–21], and may also present a viable route to modulate ion diffusion in WO$_3$ thin films in electrochromic and neuromorphic devices.

Here, we investigate the migration of different ions in WO$_3$ with *ab initio* methods, with a focus on the effect of electron polarons in the vicinity of a migrating ion and on the effect of lattice strain. We explore these effects for three technologically relevant ions of different size and charge, H$^+$, Li$^+$, and Mg$^{2+}$.

## 2. Methods

*Ab initio* calculations were performed with the VASP (Vienna Ab-initio Simulation Package) software[22–25], using the SCAN meta-GGA exchange-correlation functional[26]. To account for strongly correlated W 5d electrons, we used the DFT+U approach with $U_{W,d}$ set to 7 eV (optimized in a separate study)[27]. This ensured that polarons remained localized throughout all migration calculations to assess their impact on energy barriers. We used the VASP pseudopotentials W_sv (5p5d6s), O (2s2p), H (1s), Li(1s2s2p) and Mg(2s). We included spin polarization in the calculations and atomic coordinates were relaxed until forces were below 10 meV/Å.

In agreement with experiment[7], we chose the monoclinic structure for WO$_3$, which is stable at room temperature, as a ground state structure, and modeled the intercalation of small amounts of monovalent or divalent ions (the resulting stoichiometry amounted to $X_{0.03}WO_3$). We used $\sqrt{2} \cdot \sqrt{2} \cdot 2$ supercells with 32 W and 96 O atoms for all calculations and the Brillouin zone integration was carried out on a 3x3x2 k-point mesh for the relaxation of equilibrium structures and on a 2x2x1 k-point mesh during migration calculations. The plane-wave energy cutoff was set to 600 eV and an energy convergence criterion of 10$^{-6}$ eV was chosen. For calculations without polarons, the neutral lattice was initially relaxed with the SCAN functional and the electrons that are introduced by intercalation were removed for barrier calculations and compensated by a homogeneous background charge. For calculations with polarons, initial relaxation of structures with localized polarons was performed with the SCAN+U approach. For a proper localization of polaronic charge carriers, we introduced initial distortions in the structure by replacing a W atom with a larger Po atom. Subsequently, we reintroduced W and provided an initial magnetic moment during the first relaxation steps, thereby inducing the polaron localization on the desired site.[27] Continuous localization of polarons was checked by tracking the magnetization during relaxation steps. Migration barriers were calculated by using the Climbing-Image Nudged Elastic Band (CI-NEB) method from the VASP Transition State Tools package[28,29], providing relaxed starting structures and using five intermediate image structures between these starting structures. The artificial spring constant was set to 5 eV/Å$^2$. For biaxial strain calculations, we applied 1 % of biaxial tensile strain perpendicular to the migration path of the migrating ion. To identify the appropriate compensating strain in the third lattice direction, we varied the strain in the third lattice direction from 0 to -2.5 % in 6 steps. The appropriate compensating strain was subsequently identified with the minimum of a quadratic fit of the total energy values of the six structures for each migrating ion.

## 3. Results and Discussion

### 3.1. Structure of $WO_3$ with intercalated ions

*Ab initio* calculations of monoclinic $WO_3$ with the SCAN functional and without intercalated ions yielded lattice constants and angles in very good agreement with experimental values (see Table 1). The equilibrium position of an intercalated ion is distinctly different for the small proton and for the larger $Li^+$ and $Mg^{2+}$ ions (equilibrium structures of $WO_3$ with intercalated ions are shown in Fig. 1). A proton in stable positions is always closely attached to an oxygen atom in the $WO_3$ lattice with a short O-H covalent bond length of 0.98 Å, being in excellent agreement with earlier studies[30,31]. In contrast, $Li^+$ and $Mg^{2+}$ ions are located in the empty space that would be occupied by the A-site cation in an $ABO_3$ perovskite structure. Both ions are slightly displaced from the exact center position of the A-site, allowing for shorter Li-O and Mg-O bond lengths. The shortest Li-O and Mg-O distances are 2.00 Å and 1.98 Å, respectively.

Table 1: Lattice parameters and angles of $WO_3$ structures determined experimentally from ref.[50] and computationally (SCAN functional) in this work.

|   | experiment ($WO_3$) | calculation ($WO_3$) | calculation $H_{0.03}WO_3$ | calculation $Li_{0.03}WO_3$ | calculation $Mg_{0.03}WO_3$ |
|---|---|---|---|---|---|
| a | 7.30 Å | 7.35 Å | 7.39 Å | 7.36 Å | 7.60 Å |
| b | 7.54 Å | 7.49 Å | 7.45 Å | 7.59 Å | 7.54 Å |
| c | 7.69 Å | 7.68 Å | 7.67 Å | 7.56 Å | 7.42 Å |
| α | 90.0° | 90.0° | 89.9° | 89.9° | 90.0° |
| β | 90.9° | 91.0° | 90.9° | 91.0° | 90.8° |
| γ | 90.0° | 90.0° | 90.1° | 90.1° | 90.0° |
| Volume | 422.88 Å$^3$ | 422.11 Å$^3$ | 422.20 Å$^3$ | 422.62 Å$^3$ | 425.45 Å$^3$ |

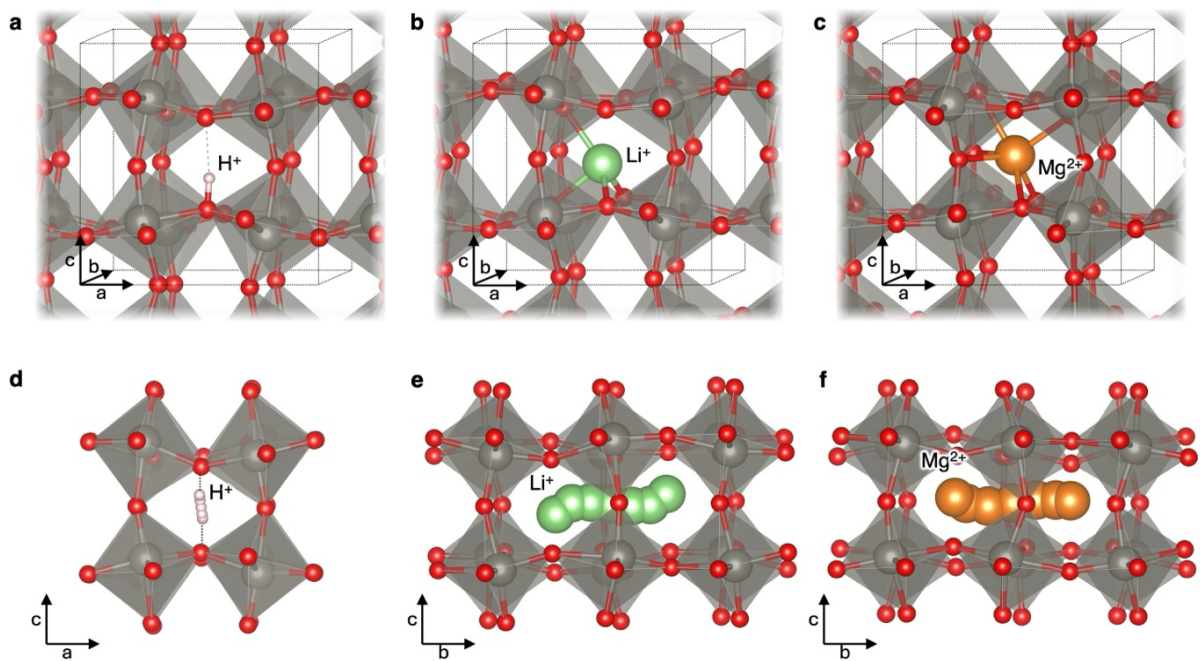

*Figure 1: Equilibrium structures of intercalated a) $H^+$, b) $Li^+$ and c) $Mg^{2+}$ ions in monoclinic $WO_3$ with the corresponding migration pathways (d-f) with five images between stable starting and end structures.*

For calculations of a neutral cell with the SCAN functional, the cell volume does not change substantially upon $H^+$ or $Li^+$ intercalation. $Mg^{2+}$ intercalation leads to a slight increase of the volume. Additionally, upon $Mg^{2+}$ intercalation, the lattice parameters experience the strongest relative changes, and the structure becomes closest to a cubic crystal structure as seen by both W-O-W angles and lattice parameters. In this study, we limit investigations to low concentrations of intercalated ions (x=0.03), as it has been previously shown that higher concentrations lead to phase transitions of monoclinic $WO_3$ (e.g. transition to a tetragonal phase for $x > 0.1$ in $Li_xWO_3$[32]), and the delocalization of the electronic carrier[33]. In addition, low ion concentrations and high resistivity are the target operating regime for the application of $WO_3$ in ECRAM devices in hardware AI accelerators[34].

## 3.2 Ion migration pathways and barriers

To assess the mobility of $H^+$, $Li^+$ and $Mg^{2+}$ ions, we sampled different migration pathways. Given that a proton is closely bound to an O atom, different possible migration paths need to be investigated, with the most relevant processes being interoctahedral transfer of $H^+$ to the opposite oxygen atom (second nearest neighbor), intraoctahedral transfer to an oxygen atom with a shared W atom (first nearest neighbor) and rotational movements around oxygen atoms. In the related $ABO_3$ perovskite structure, it was shown e.g. for $NaTaO_3$ or $BaFeO_3$ that protons preferentially transfer intraoctahedrally between nearest neighbor oxygen atoms with a shared B-site corner[35–37]. For doped $BaZrO_3$, it was found that the predominant pathways for proton transfer are also intraoctahedral, however, local distortions induced by dopants unlock interoctahedral transfer[38]. We find that for $WO_3$, intraoctahedral transfer is not the most favorable pathway and leads to high migration barriers (>0.6 eV, see Supporting Information S.I.1). Instead, protons prefer to transfer interoctahedrally between opposite oxygen atoms, leading to significantly lower migration barriers (see Fig. 1a and 2a). We screened various pathways for interoctahedral proton transfer in monoclinic $WO_3$ and found comparatively low energy barriers between 0.26 and 0.38 eV, depending on the initial distance between the two oxygen atoms (see Fig. 2b). In contrast to the A-site-filled $ABO_3$ perovskite structure, in $WO_3$ the A-site is vacant, leading to increased lattice flexibility and distortion that allows opposite oxygen atoms to move towards each other during interoctahedral proton transfer. In addition, the presence of no additional cation along the pathway reduces the electrostatic repulsion of the proton along this migration path. On the other hand, we suspect that the highly positively charged W cation in the center of $WO_6$ octahedra prevents the positively charged proton from travelling intraoctahedrally between corner-shared oxygen atoms. It is noteworthy that interoctahedral proton transfer and rotation pathways lead to percolating diffusion for protons through $WO_3$.

For bigger ions, i.e. $Li^+$ and $Mg^{2+}$, which reside in the center of the empty space between the $WO_6$ octahedra, the most likely migration pathways resemble A-site cation migration processes in a perovskite structure[39,40]. Here the cation traverses through a window of 4 oxygen anions and 4 B-site cations (in this case W, see Fig. 1b and c). In the $WO_3$ structure, this migration process is possible in the [100], the [010], and the [001] direction, leading to a range of different possible energy barriers for $Li^+$ and $Mg^{2+}$ (0.55 - 0.63 eV, and 1.34 - 2.01 eV for $Li^+$ and $Mg^{2+}$, respectively, see Fig. 2b and c), due to the anisotropy of the monoclinic crystal structure (the single barriers for different migration directions are shown in the Supporting Information S.I.2).

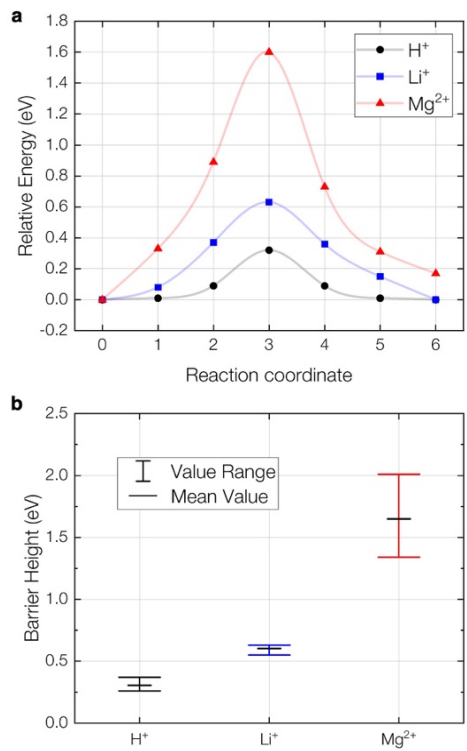

*Figure 2: a) Energy landscape of representative ion migration pathways for $H^+$, $Li^+$ and $Mg^{2+}$ in monoclinic $WO_3$, as shown in Fig. 1 d-f). b) Range of values for energy barriers of different migration directions (different O-O transfers for $H^+$, and different migration directions for $Li^+$ and $Mg^{2+}$)*

Importantly, calculations so far have considered exclusively the migration of the isolated ion in the monoclinic $WO_3$ lattice without any additional electronic charge carriers.[20,41] In the following, we will therefore explore the impact of electron polarons in the vicinity of intercalated ions on the energy landscape during ion migration.

## 3.3 Polaron effects on ion migration processes

Electrons that are introduced by the intercalation of low concentrations of $H^+$, $Li^+$ and $Mg^{2+}$ ions are usually localized in polaronic form in the $WO_3$ lattice[7,17,18]. Experimental studies generally report localization in the form of small to intermediate polarons[18,41,42] (e.g. with an estimated radius of 5-6 Å[18]). This picture is also substantiated by calculated polaron binding energies around 0.3 eV[27,41,43,44]. The exact size of electron polarons in DFT calculations depends on computational details. With expensive hybrid functionals, a Hartree-Fock exchange parameter of 0.5-0.6 was found to reproduce experimental values of the localized charge density on the central tungsten atom (~50%)[45]. In this work, the use of the SCAN+U method results in around 75% of charge localized on the central tungsten atom, slightly underestimating the electron polaron size.

To assess the effect of electron polarons on ion migration, we calculated the ion migration barriers for the energetically most favorable polaron-proton configurations. To estimate upper bounds for migration barriers (maximal association), we limited calculations to configurations, where the distance between the ion and the polaron increases during the migration process, i.e. where the migrating ion "escapes" the polaron. The most favorable ion-polaron configurations and the corresponding ion migration pathways are shown in Fig. 3 a-c.

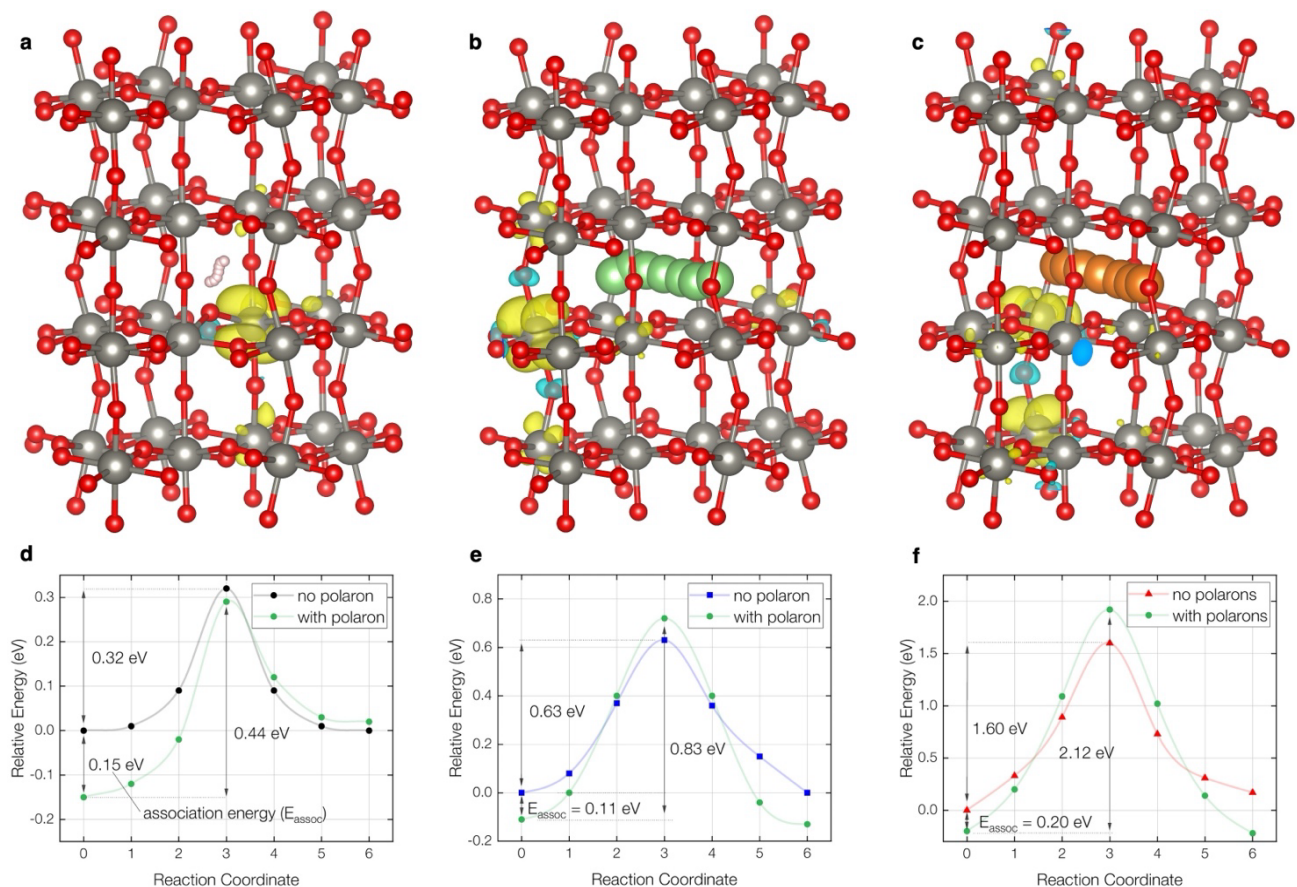

*Figure 3: Energetically most favorable configurations for the localized small polaron in the vicinity of an intercalated ion, a) $H^+$, b) $Li^+$ and c) $Mg^{2+}$. d-f) Energy landscape during migration without and with an electron polaron in the configurations shown in a-c), respectively.*

Upon calculation and comparison of the migration barriers in the presence of polarons, we identified two important effects that are induced by the polaron and that affect the migration of ions in $WO_3$. First, the polaron and the mobile ion can form an associated defect complex when brought close together, primarily due to electrostatic interaction, and an energy penalty of this association must be overcome when moving the ion away from the polaron. The energy gain caused by ion-polaron association ($E_{assoc}$) was estimated according to $E_{assoc}=E_{ion+polaron}-E_{distant\ ion+polaron}$ where the ion was placed next to the polaron ($E_{ion+polaron}$) and at an equivalent lattice site at least 10 Å away from the polaron ($E_{distant\ ion+polaron}$), both calculated in a 2x2x2 supercell. Second, polarons induce changes of the local lattice geometry that affect interatomic distances and angles, and thus the migration of mobile ions.

For protons, the dominant effect of the electron polaron is the association, and the resulting association energy is 0.15 eV. Evaluating the total barrier height shows that the escaping ion needs to overcome a barrier of 0.44 eV, which is 0.12 eV higher than without a polaron (see Fig. 3 d). Interestingly, this value is also much closer to experimentally determined activation energies of proton migration[9,13], suggesting that in most realistic experimental settings, it is essential to consider the effect of polaronic charge carriers on the migration of protons in $WO_3$.

In the case of $Li^+$ and $Mg^{2+}$, the association energies were 0.11 eV and 0.20 eV, respectively. In addition, here, a second effect of the polaron on the migration barrier arises from lattice distortions that are induced by the presence of the polaron. The spacing between the two closest oxygen atoms that the $Li^+$ ion needs to pass decreases from 3.18 Å to 3.10 Å (top and bottom oxygens in Fig. 4 b). Quantitatively, the overall migration barrier of the $Li^+$ ion increases by 0.2 eV upon introduction of a small polaron. For $Mg^{2+}$ ions, two polarons induce changes in the lattice geometry that again reduce the free space for the $Mg^{2+}$ ion in the transition state. For $Mg^{2+}$, the O-O spacing is reduced from 3.21 Å to 3.03 Å and the two effects account for an overall increase in the migration barrier height by 0.42 eV. The combination of our computational method (overestimation of localization) and our choice of the lowest-energy proton-polaron configurations makes these results upper-bound estimates for the effect of electron polarons on ion migration barriers, but ultimately, their presence increases migration barriers in all cases.

### 3.4. Strain effects on ion migration in $WO_3$

Lattice strain can be introduced in the form of elastic epitaxial strain at lattice-mismatching interfaces [19,21], or in the form of lattice and density disorder[46]. To assess the effect of lattice strain on the energy barriers of ion migration processes, we selected one representative migration process and applied isotropic tensile and compressive strain to the crystal lattice (-2%, -1%, 1% and 2%). Interestingly, strain has contrary effects on protons and on larger ions (see Fig. 4 a). For protons, compressive strain decreases the barrier for proton transfer between two opposite oxygen atoms, and tensile strain increases this barrier. More specifically, 1% compressive strain reduces the energy barrier by 0.06 eV from 0.32 to 0.26 eV. For $Li^+$ and $Mg^{2+}$, the opposite effect is observed; tensile strain decreases the migration barrier. For $Li^+$, 1% tensile strain reduces the energy barrier by 0.23 eV from 0.55 to 0.32 eV; and for $Mg^{2+}$, 1% tensile strain reduces the barrier by 0.46 eV from 1.60 to 1.14 eV. This difference in strain response of the proton versus that of $Li^+$ and $Mg^{2+}$ is caused by the fundamental difference in the migration mechanism. In the case of proton migration, the proton is embedded in the oxygen ion electronic shell and transfers between a donor and an acceptor oxygen ion. Reducing the distance between these oxygen atoms by applying compressive strain leads to a decrease in the energy barrier of this transfer step (depicted in Fig. 4 b). This is consistent with the results of Hoedl et al., who showed that the barrier for proton transfer in $BaFeO_3$ is strongly dependent on the initial distance between donor and acceptor oxygen ions (transfer along compressed octahedral edges is significantly easier)[47]. For larger ions that are initially located in the empty interstitial space of $WO_3$, the transition state lies in a bottleneck containing 4 oxygen and 4 W atoms, and the migrating ion repels the two closest oxygen atoms during this process. Therefore, applying tensile strain increases the space available for the migrating ion and reduces the energy barrier[48,49] (depicted in Fig. 4 b). We believe that this is also the underlying reason for the different slopes observed between $Li^+$ and $Mg^{2+}$ migration, where tensile strain has a stronger effect on the larger $Mg^{2+}$ ion.

In addition, we investigated the impact of strain also for ion migration in the presence of polaronic charge carriers, in the same range as above, -2% compressive to +2% tensile strain. The resulting values for the migration barriers are again shown in Fig. 4 a. The calculations confirm our expectations that the general effects of strain on ion migration in $WO_3$ are preserved in the presence of small polarons. While compressive strain facilitates proton migration, it increases the migration barrier for larger ions; and vice versa for tensile strain. The increase of the migration barrier by introduction of an electron polaron that is observed for all ions shows a weak dependence on the strain state, generally, the barrier increase is less pronounced for lattices under tensile strain. We believe that this is related to reduced lattice distortions that are introduced by electron polarons in the larger, expanded lattice.

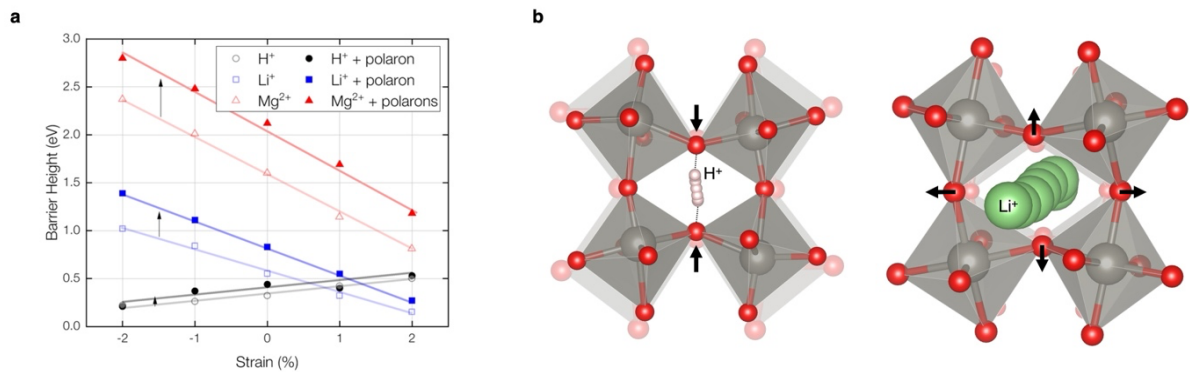

Figure 4: a) Energy barriers of representative migration pathways for $H^+$, $Li^+$ and $Mg^{2+}$ ions in monoclinic $WO_3$ upon application of tensile and compressive lattice strain and combined effects of strain and polarons on the migration barrier height. b) schematics of the effects of strain on ion migration in $WO_3$.

We also assessed the effect of biaxial strain on migration barriers of $H^+$, $Li^+$ and $Mg^{2+}$ ions in $WO_3$ as this is the expected scenario of thin epitaxial films on lattice-mismatching substrates. We applied 1% biaxial tensile strain perpendicular to the migration direction, leading to a compensating compressive strain along the migration direction (-1.3% for $H^+$, -0.13% for $Li^+$, and -0.65% for $Mg^{2+}$). For $H^+$ migration, this shortens the distance between the acceptor and donor O atoms during a proton transfer. For $Li^+$ and $Mg^{2+}$ ions, this strain leads to a widening of the oxygen bottleneck that the ion has to pass through in the transition state. Hence, according to the above considerations, for all ions, compressive strain along the migration direction should decrease the migration barrier. Our calculations confirm this trend, with the magnitude of the achievable change in migration barrier being smaller than for isotropic strain. As shown in Fig. 5, in the case of protons, the migration barrier was decreased only slightly by 0.02 eV from 0.32 to 0.30 eV. For Li$^+$ ions, the barrier was reduced by 0.25 eV from 0.63 to 0.38 eV and by 0.28 eV from 1.60 to 1.32 eV for Mg$^{2+}$ ions. Considering these results, biaxial strain could be leveraged to accelerate ion migration in a desired direction (e.g. normal to the substrate), in particular for larger ions.

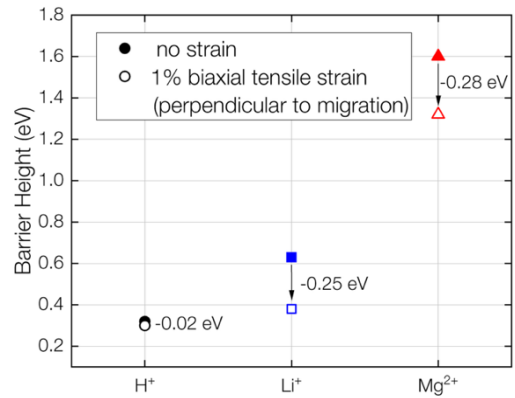

*Figure 5: Changes in the energy barriers of ion migration upon introduction of 1 % biaxial tensile strain perpendicular to the migration direction.*

## 4. Conclusions

In this study, we investigated the effects of small polarons and lattice strain on the migration of H$^+$, Li$^+$ and Mg$^{2+}$ ions in monoclinic WO$_3$, a process of high relevance for electrochromic applications, energy storage, and neuromorphic computing devices. We found that protons are closely attached to oxygen atoms in the WO$_3$ lattice and migrate via transfer from donor to acceptor oxygen atoms along the [100], [010] and [001] directions, different from proton transfer between nearest neighbor oxygen atoms in the related ABO$_3$ perovskite structure. In contrast, Li$^+$ and Mg$^{2+}$ ions are centrally located in the empty A-site space of the WO$_3$ lattice and migrate between these empty sites along the [100], [010] and [001] directions. Without the presence of polaronic charge carriers, H$^+$ exhibits the lowest average migration barrier of ~0.3 eV, followed by Li$^+$ with ~0.6 eV and Mg$^{2+}$ with ~1.6 eV. Upon the introduction of small polarons as compensating electronic species, migration barriers are changed significantly. In the case of H$^+$, strong association between the ion and the small polaron increases the migration barrier height by 0.15 eV. The resulting migration barrier height of ~0.45 eV is in excellent agreement with experimental proton diffusion activation energies, suggesting that polaron effects must not be neglected when discussing proton migration in WO$_3$. In case of Li$^+$ and Mg$^{2+}$ ions, introduction of polarons also leads to increased migration barriers, caused both by ion-polaron association, and lattice distortions that are introduced by the polaron, leading to a less favorable transition state geometry. Applying lattice strain can lead to substantial changes of the migration barriers, and protons respond fundamentally differently to the applied lattice strain in comparison to larger ions. Proton migration exhibits a lower barrier when the lattice is compressed, i.e. when the the donor and acceptor O atoms are brought closer. In contrast, the migration of Li$^+$ and Mg$^{2+}$ ions benefits from tensile lattice strain, as it increases the available space around the ions in the transition state. Combining the effects of strain and polarons, we confirm that the effects of lattice strain on ion migration processes is conserved in the presence of polarons. Lastly, we show that biaxial strain that is compressive along the migration direction reduces the migration barrier of H$^+$, Li$^+$ and Mg$^{2+}$ ions in WO$_3$, underlining the technological applicability of the results, for example, for tailored migration properties in thin films.

## Acknowledgements

M.S. acknowledges the support of the Max Kade Foundation through a Max Kade fellowship. This research has been partially supported by SUPREME Center within the Semiconductor Research Corporation's JUMP 2.0 program (Pillar ID 3137.036). The authors acknowledge supercomputer resources provided by the Frontera computing project DMR20012 at the Texas Advanced Computing Center, the MIT Engaging Cluster, and MIT Supercloud.

# Supporting Information: Polaron and Strain Effects on Ion Migration in WO$_3$

Matthäus Siebenhofer, Pjotrs Žguns, Bilge Yildiz

S.I.1. Different proton migration pathways

WO$_3$ and ABO$_3$ perovskites exhibit a largely similar lattice structure consisting of oxygen octahedra with cations in their center positions. The critical difference is that perovskites, in addition, contain A-site cations between these octahedra, which are empty sites in WO$_3$. Given their structural similarity, we would expect that the energetically favorable proton migration pathways are largely similar in the two material classes. In perovskites, it has been shown that the favorable migration pathways for protons are between nearest neighbor oxygens, i.e. along the edges of the BO$_6$ octahedra[35–37]. To evaluate the most favorable migration step in WO$_3$, we compared the following two possible migration steps (see also Figure S. 1):

i) Direct proton transfer between two next-nearest neighbor oxygen atoms, with the transfer path connecting the corners of two adjacent octahedra (interoctahedral transfer).
ii) Proton transfer via a nearest neighbor oxygen atom, following the proton migration paths that have been suggested for perovskites (intraoctahedral transfer)

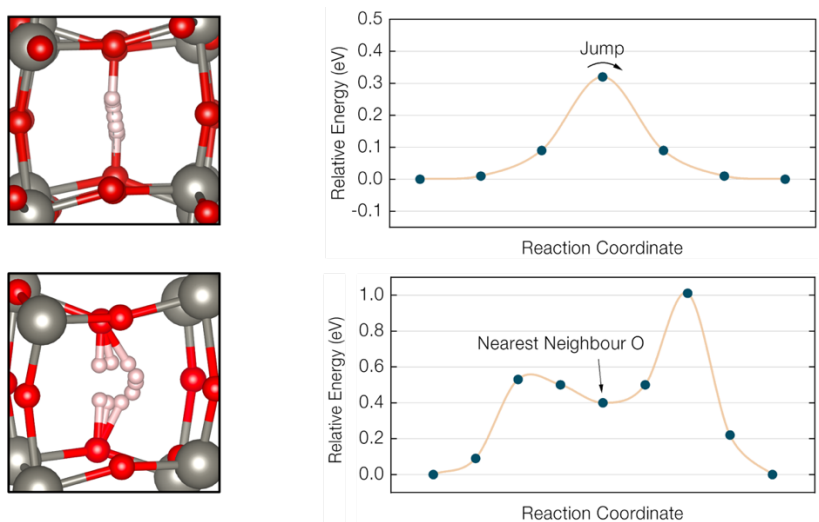

*Figure S. 1: Visualizations of two different proton migration pathways between next-nearest neighbor oxygen atoms in WO$_3$ and the corresponding migration barriers resulting from CI-NEB calculations.*

Surprisingly, the calculations show that in WO$_3$, it is not favorable for a proton to transfer intraoctrahedrally between nearest neighbor oxygen atoms, in contrast to proton migration in perovskites. Instead, protons transfer interoctahedrally between next nearest neighbor oxygen atoms straight across the empty space in between the two oxygen atoms. The difference in migration barrier is significant. While the (polaron-free) migration barrier for interoctahedral transfer is estimated with 0.32 eV, the migration barrier for intraoctahedral transfer amounts to ~1eV, rendering these movements essentially irrelevant at room temperature.

S.I.2. Variation of energy barriers for ion migration

Due to the not fully symmetric and cubic lattice geometry of monoclinic $WO_3$, ion migration processes along different lattice directions are generally not equivalent. This leads to a certain variation of energy barriers that is shown in Fig. 2 b) in the main manuscript. Here, we show the corresponding energy landscapes for the surveyed migration processes. The energy spread is smallest for $Li^+$ migration (ca. 0.1 eV) and largest for $Mg^{2+}$ migration (ca. 0.6 eV).

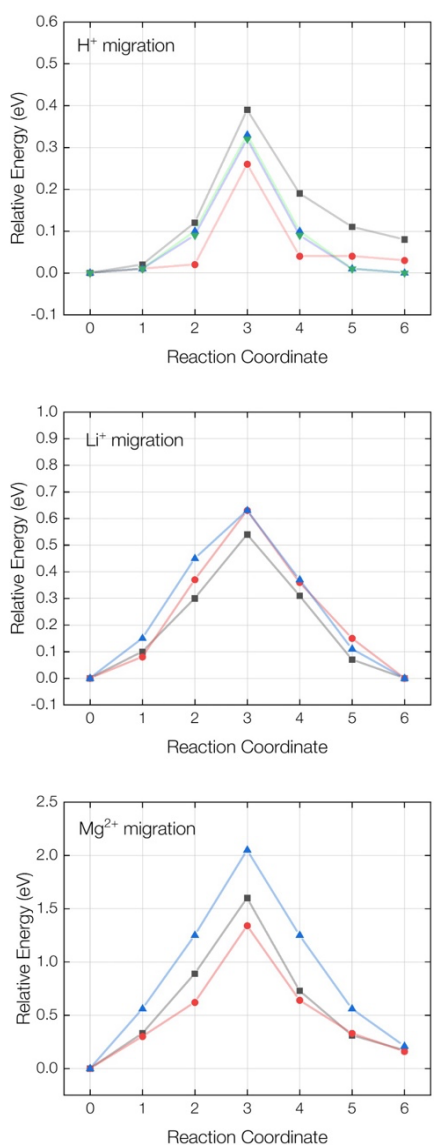

*Figure S. 2: Energy landscapes for various ion migration processes in $WO_3$.*